\documentclass[twocolumn,prd,nofootinbib]{revtex4}

\usepackage{graphicx}

\usepackage{epsf}
\usepackage{dcolumn}

\newcommand{\be}{\begin{equation}}
\newcommand{\ee}{\end{equation}}

\def\4he{$^4$He}
\def\3he{$^3$He}
\def\7li{$^7$Li}

\newcommand\la{\lower0.6ex\vbox{\hbox{\ensuremath{\buildrel{\textstyle<}\over{\sim}\ }}}}
\newcommand\ga{\lower0.6ex\vbox{\hbox{\ensuremath{\buildrel{\textstyle>}\over{\sim}\ }}}}
\newcommand{\roughly}[1]%
    {{\mathrel{\raise.3ex\hbox{$#1$\kern-.75em\lower1ex\hbox{$\sim$}}}}}
\newcommand{\BM}[1]{{\mbox{\boldmath{$#1$}}}}

\def\lsim{\mathrel{\raise.3ex\hbox{$<$\kern-.75em\lower1ex\hbox{$\sim$}}}}
\def\gsim{\mathrel{\raise.3ex\hbox{$>$\kern-.75em\lower1ex\hbox{$\sim$}}}}

\begin{document}

\title{PAMELA and dark matter}
\author{V.~Barger$^{1}$, W.-Y. Keung$^2$, D.~Marfatia$^{3}$, G. Shaughnessy$^{1,4,5}$}
\affiliation{$^1$Department of Physics, University of Wisconsin, Madison, WI 53706}
\affiliation{$^2$Department of Physics, University of Illinois, Chicago, IL 60607}
\affiliation{$^3$Department of Physics and Astronomy, University of Kansas, Lawrence, KS 66045}
\affiliation{$^4$Department of Physics and Astronomy, Northwestern University, Evanston, IL 60208}
\affiliation{$^5$High Energy Physics Division, Argonne National Laboratory, Argonne, IL 60439}

\begin{abstract}

Assuming that the positron excess in PAMELA satellite data is a consequence of annihilations of cold dark
matter, we consider from a model-independent perspective
if the data show a preference for the spin of dark matter. We then perform a general analysis of
 annihilations into two-body states to determine what weighted combination of channels best
describes the data.

\end{abstract}
\maketitle

{\bf{Introduction.}} 
One of the unsolved problems in cosmology and particle physics is the nature of dark matter
(DM) which accounts for about 20\% of the energy density of the universe.
Particle physics models typically relate discrete symmetries with
the existence of a stable cold DM candidate. 
A variety of such models have been suggested that provide viable explanations of the DM.
Weakly interacting massive particles (WIMPs) can be either of integer or noninteger spin. 
The classic case of supersymmetry (SUSY) has a spin-1/2 neutralino as 
dark matter whereas extra dimensional models
and collective symmetry breaking models have spin-1 dark matter. 
Specific realizations are the minimal Universal Extra Dimensions (mUED)~\cite{Appelquist:2000nn} 
and Little Higgs with T-parity (LHT)~\cite{Cheng:2003ju} models. Spin-0 dark matter
is possible in models with an additional singlet in the scalar sector of the Standard
Model~\cite{xSM}.

Recent evidence for a positron excess in the Payload for Matter Antimatter Exploration and Light-nuclei Astrophysics (PAMELA) 
data~\cite{pamela} spurs attention to WIMPs
whose annihilations in the galactic halo can explain 
an excess over backgrounds~\cite{Bergstrom:2008gr,Cirelli:2008jk}.  
PAMELA data presented thus far show a turn-up in the energy 
spectrum at about 10 GeV and a steady rise up to 100 GeV
with no fall-off in that dataset.{\footnote{We do not consider the excess in the $e^++e^-$ energy
spectrum between 500 and 800~GeV seen by the PPB-BETS balloon experiment~\cite{bets}.}} 
The shape of the spectrum bears directly on the annihilation
mechanism. Spin-1 DM annihilations directly 
into $e^+e^-$ produce a line spectrum, whereas 
spin-1/2 Majorana DM will give a continuum spectrum from secondary 
decays of weak bosons, quarks and leptons. 

Our goals are to study in a model-independent approach if the PAMELA excess provides hints about the spin of the DM particle
 and what annihilation channels are favored by the data.
 We do not subscribe to any specific particle physics model, but comment 
on models where appropriate. We also do not require that the measured relic abundance be reproduced since this is
highly model-dependent. Moreover, the total energy density in dark matter may be comprised
of several components, so only an upper bound need be imposed on the energy density of a particular DM
particle. The nature of our analysis precludes us from
 making projections for signatures at IceCube, the Fermi Gamma-ray Space Telescope, direct
detection experiments and the Large Hadron Collider, all of which are interesting in their 
own right.

\textbf{Modelling the positron signal and background.}
The positron background expected primarily from supernovae and from collisions 
of cosmic ray 
protons and nuclei on the interstellar medium is simulated in
Ref.~\cite{Moskalenko:1997gh}. The results of the 
simulation have the convenient 
parameterization~\cite{Baltz:1998xv}, 
$\Phi^{bkg}_{e^+} = 4.5 E^{0.7}/(1+650 E^{2.3}+1500 E^{4.2})$, with the energy
of the positron $E$ in GeV. Since we present 
our results as a positron fraction $\Phi_{e^+}/(\Phi_{e^+}+\Phi_{e^-})$ which 
allows for cancellations of systematic uncertainties and the effects of 
solar activity, we also need the electron background which is analogously 
parameterized by~\cite{Cirelli:2008id}
$\Phi^{bkg}_{e^-} = 0.16 E^{-1.1}/(1+11E^{0.9}+3.2 E^{2.15})+0.7 E^{0.7}/(1+110 E^{1.5}+580 E^{4.2})$. Solar modulations arise from the phase of the
solar cycle and from the opposite charges of electrons and positrons. 
Without charge sign bias, the positron ratio is independent of solar 
activity. However, since PAMELA data show evidence of charge sign dependence 
for positron energies below 10~GeV, we only analyze data above 10~GeV.

Positrons produced in DM annihilations propagate through the interstellar medium
to the earth and as a consequence suffer absorption effects that broaden the 
positron spectrum to lower energies. 
We estimate the primary positron flux from dark matter annihilations according
to the prescription of 
Refs.~\cite{Hisano:2005ec,Delahaye:2007fr,Cirelli:2008id}. 
Here we briefly describe the procedure 
and refer the reader to Refs.~\cite{Delahaye:2007fr,Cirelli:2008id} for details. 

The positron number 
density per unit energy is governed by the diffusion-loss equation with 
diffusion coefficient $K(E)=K_0 (E/\rm{GeV})^{\delta}$ which describes
propagation through turbulent magnetic fields, and is taken to be independent
of position. The equation also accounts for energy losses
through synchrotron radiation and inverse Compton scattering on the cosmic 
microwave background and infrared galactic starlight. The diffusion zone
in which the diffusion-loss equation is applicable is modelled as a cylinder
of height $2L$ and radius 20 kpc that straddles the galactic plane in which
most cosmic ray interactions take place. The
positron number density is assumed to vanish on the surface of the cylinder,
since outside the diffusion zone the positrons propagate freely and 
escape into the intergalactic medium. The source of the positrons due to
DM annhilations depends on the DM density profile and on the
annihilation cross section. For the former, 
we adopt a cored isothermal halo profile~\cite{Bahcall:1980fb}.

The normalization $K_0$ and the spectral index $\delta$ of the diffusion 
coefficient, and $L$ can all be selected to be consistent with the measured
boron to carbon ratio in cosmic rays~\cite{Maurin:2001sj}. 
We consider three sets of these
parameters, ``Min'', ``Med'' and ``Max'', of
Ref.~\cite{Delahaye:2007fr} and reproduce them in Table~\ref{table1}. 
The Med set has values of $K_0$, 
$\delta$ and $L$ which best fit the measured boron to carbon ratio. The Min and 
Max sets minimize and maximize the positron fluxes above about 10 GeV. 
Needless to say, the Min and Max sets are only representative, since
the positron flux depends on the mass of the DM particle $M_{DM}$ and 
on the annihilation channel once
the halo profile is selected.

\begin{table}[t]
\begin{tabular}{lcccc}
\hline
Model & $\delta$ & \ $K_0$ (kpc$^2$/Myr)\ & \ $L$ (kpc)  \\
\hline
Min & 0.55 & 0.00595 & 1 \\
Med  & 0.70 & 0.0112 & 4 \\
Max & 0.46 & 0.0765 & 15 \\
\hline
\end{tabular}
\caption{Three sets of parameters describing cosmic ray propagation~\cite{Delahaye:2007fr}. The Med set 
is the best-fit to the measured boron to carbon ratio. The Min and Max sets minimize and maximize
the positron fluxes above 10~GeV, respectively. }
\label{table1}
\end{table}

Assuming steady-state conditions, a semi-analytic expression for 
the primary positron flux has been 
obtained~\cite{Hisano:2005ec,Delahaye:2007fr}. The result depends on a
so-called ``halo function'' which encodes the physics of cosmic ray 
propagation. We employ numerical fit functions~\cite{Cirelli:2008id} 
for the halo functions pertinent to the isothermal profile with the Min, Med and 
Max propagation parameter sets. We allow for the possibility of high density 
substructures in the dark matter halo 
that enhance the positron flux by an energy-independent ``boost factor'' $B$. 
Note that $N$-body simulations suggest that $B$ can not be larger 
than about 10 and
may be energy-dependent~\cite{Lavalle:1900wn}.

\textbf{Dark matter annihilations and spin.}
To begin with, we assume $M_{DM}$ is smaller than the top quark mass. 
This choice is dictated by our interest in model-independence. (We shall see later, by extending 
the range of $M_{DM}$, that the positron data typically select DM lighter than the top quark for the Med set). 
If the $t\bar{t}$ channel were open, the relative contributions of different channels
 to the positron spectrum would depend on
the details of the Higgs sector. 

Since the PAMELA data show a sharp rise, we initially only consider
positrons from a $e^+$ line spectrum or from the 
two-body decays of pair-produced weak bosons at the source. 
Specifically, we study the spectra from direct production, $DM DM \rightarrow e^+e^-$
(which produces a positron line close to $M_{DM}$), and 
from secondary production from the process $DM DM \rightarrow W^+W^-$.{\footnote{Although 
each $Z$ in a $Z$ pair produces a positron, since 
${\sigma(DM DM\to W^+W^-)/ \sigma(DM DM \to ZZ)} \approx 2$ in the high-energy limit, and since
the leptonic branching fraction for $W$s is three times as much as for $Z$s, $W$ pairs
produce three times as many positrons as $Z$ pairs with almost identical distributions. }}
Concrete examples of direct annihilation into $e^+e^-$ are found in mUED and LHT in which
spin-1 DM annhilate by exchange of an odd-parity fermion~\cite{Servant:2002aq,Birkedal:2006fz}. 
Direct annihilation also occurs for hidden/mirror Dirac fermions and sterile neutrinos. The 
latter constitute warm DM which is not relevant to our study of nonrelativistic DM. 
If DM is a Majorana fermion, helicity suppression prevents the direct production of $e^+e^-$.
For scalar DM the amplitude for static annihilation into light fermions vanishes~\cite{Barger:2008qd}. 
Since the production of $W$ pairs is 
spin-dependent, we further classify the positron spectra according to whether the $W$ bosons are
longitudinally polarized or tranversely polarized.

The normalized distributions for the $e^+$ energy are
\begin{eqnarray}
 f_{TT}(x)&=&3{ \beta_W^2 + (1-x)^2 \over 8\beta_W^3}\\
 f_{LL}(x)&=&3{ \beta_W^2 - (1-x)^2 \over 4\beta_W^3}\,,
\end{eqnarray}
for the $WW$ transverse ($TT$) modes and longitudinal ($LL$) mode, respectively, where
$\beta_W^2=1-m_W^2/M_{DM}^2$ and $x=2E_{e^+}/M_{DM} 
{\mathrel{\raise.3ex\hbox{$<$\kern-.75em\lower1ex\hbox{$>$}}}}1\pm\beta_W$.
In general, if the $W^+W^-$ channel has both $TT$ and $LL$ contributions of
relative weights $a$ and $b$, then the resultant distribution 
which combines two $TT$ modes and one $LL$ mode is
$ (a f_{LL}(x) +  b f_{TT}(x))/ (a+b)$. 

\begin{table}
\begin{tabular}{lcccc}
\hline
spin\ \ & \ \ \ $s$-channel \ \ \ \   & \ \ \ $t,u$-channel \ \ \ \ & $t,u$-channel  \\
\ \    & \ \ \ Higgs   \ \ \ \    & \ \ \    fermion \ \ \ \  &   boson \\
\hline
0 & LL, TT & X & LL \\
${1\over 2}$  & 0 & TT & X \\
1 &LL, TT & X & LL, TT \\
\hline
\end{tabular}
\caption{Polarizations of $W$ pairs produced by static annihilations $DM DM  \to W^+W^-$ depend
on the spin of the DM particle. ``LL'' and ``TT'' indicate that the $W$ bosons
are longitudinally and transversely polarized, respectively. ``X'' indicates
 that there is no contribution
at the tree-level, and ``0'' indicates that the amplitude vanishes in the static limit. 
Note that Dirac fermion DM also has contributions from $s$-channel 
$Z$-exchange.}
\label{table2}
\end{table}

In Table~\ref{table2}, we categorize the polarization modes of the $W$ pair according to 
the spin of the DM particle. While fermionic DM can not annihilate into $W$ pairs
via $s$-channel Higgs exchange in the static limit, spin-0 and spin-1 DM annhilations
 (with relative weights \mbox{$a=(1+\beta_W^2)^2$} and \mbox{$b=2 (1-\beta_W^2)^2$)} give the distribution,
\begin{equation} {1\over N} 
{dN\over dx}={3[1+\beta_W^4-2(1-x)^2] \over 2\beta_W(3-2\beta_W^2+3\beta_W^4)}\,.
\end{equation}
Note that as $\beta_W\rightarrow 1$, the longitudinal mode dominates. The DM particle in 
both mUED and LHT can annihilate via $s$-channel Higgs exchange. 
Whether these models produce line or continuum
spectra or both depends on specific realizations.

\begin{figure}[t]
\vskip 0.2in
\mbox{\includegraphics[width=3.25in]{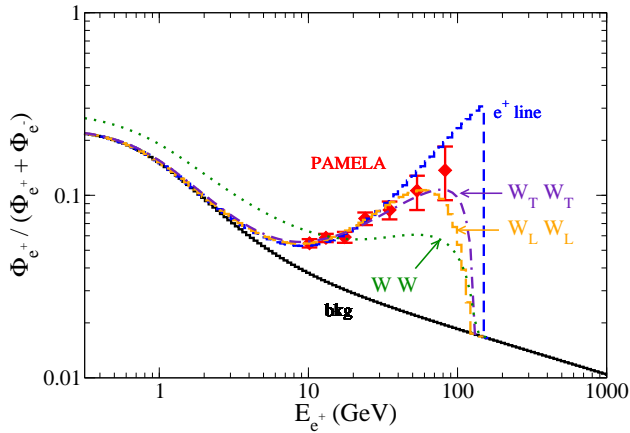}}                                      
\caption[]{Annhilations of  DM directly into $e^+e^-$ give the $e^+$ line at about 
$M_{DM}=150$ GeV. 
The secondary positron spectrum from decays into $e^+\nu$ of longitudinal (transverse) $W$ bosons                               
is labeled $W_L W_L$ ($W_T W_T$); the soft component of the spectra are neglected for illustration.
Including the soft component (with spin-correlations neglected) results in a much softer spectrum labelled
$WW$ that does not fit the PAMELA data above 10~GeV. 
The solid curve is the expected background. The Med set of propagation parameters is
used with a cored isothermal profile for the DM halo.
\label{fig:1}}
\end{figure}

Fermionic DM annhilations via $t$- or $u$- exchange of a fermion
give only $TT$ modes. The positron spectrum is then simply $f_{TT}(x)$~\cite{Barger:2006gw}. 
SUSY provides the common example of neutralinos that annihilate
dominantly by $t$- and $u$-channel chargino exchange.

\begin{figure}[ht]
\vskip 0.2in
\mbox{\includegraphics[width=3.25in]{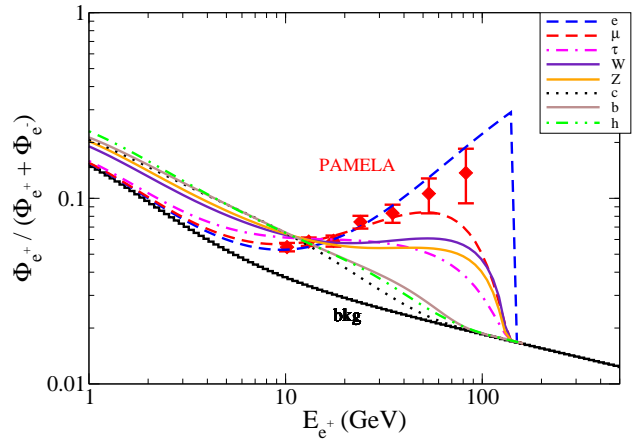}}
\caption[]{Positron fraction from DM annihilation into $e^+e^-$, $\mu^+\mu^-$, $\tau^+\tau^-$, $W^+W^-$, $ZZ$, 
$c\bar{c}$, $b\bar{b}$, and $hh$, with a Standard Model Higgs boson 
$h$ of mass 120~GeV for the Med set of propagation parameters. We have assumed that the DM annihilates into each mode 
with a 100\% branching fraction. Accounting for the smaller boost factor, the $e^+e^-$ mode is somewhat 
preferred; see Table~\ref{table3}. The Max set yields spectra very similar to the Med set.
\label{fig:2}}
\end{figure}
                                                                                                                                                                                             
\begin{figure}[ht]
\vskip 0.2in
\mbox{\includegraphics[width=3.25in]{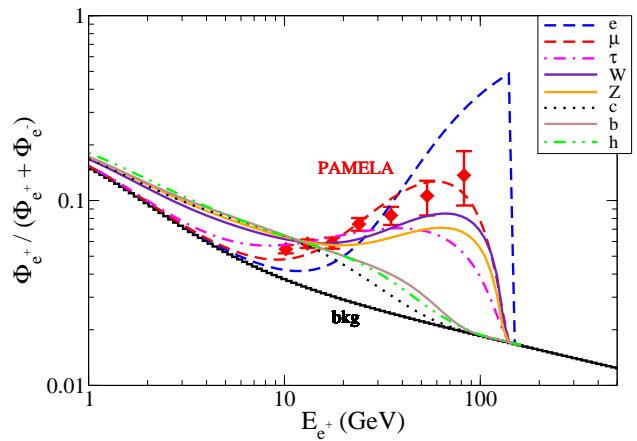}}
\caption[]{Same as Fig.~\ref{fig:2}, but for the Min set of propagation parameters. The $\mu^+\mu^-$, $\tau^+\tau^-$
and $W^+W^-$ modes are preferred; see Table~\ref{table3}. 
\label{fig:3}}
\end{figure}

\textbf{Analysis.} 
In Fig.~\ref{fig:1}, we show spectral distributions of positrons produced in annhilations
of a DM particle of mass 150 GeV that fit the PAMELA excess. 
We have assumed that when direct production occurs, annihilations into $W^+W^-$ are negligibly small. 
For the $W^+W^-$ channels, only the hard spectra from $W^+ \to e^+ \nu$ are shown. The soft
component of the spectra from the $W$ other decay modes is neglected to emphasize the
small difference between the hard spectra from $W_LW_L$ and $W_TW_T$. 
In what follows, we disregard the effects of spin-correlations. Including the soft component, we find
annihilations dominantly into $W^+W^-$ do not provide a satisfactory spectrum. 


We now enlarge the scope of our study by allowing $M_{DM}$ to be as large as 1~TeV and 
allowing arbitrary weights for the following annihilation modes:
$e^+e^-$, $\mu^+\mu^-$, $\tau^+\tau^-$, $W^+W^-$, $ZZ$, $c\bar{c}$, $b\bar{b}$, $t\bar{t}$ and $hh$, with a Higgs boson 
$h$ of mass 120~GeV which will decay primarily into $b$ and $\tau$. Annihilations into $Zh$ can be accounted for
by the average of the $ZZ$ and $hh$ channels. The subsequent decays were computed using micrOMEGAs~\cite{micromegas}. 
We denote the weights by $f_{xy}$, where for example,
$f_{e^+e^-}$ is the weight of the $e^+e^-$ channel.
In Figs.~\ref{fig:2} and~\ref{fig:3}, we show the positron fraction from each of these channels (except $t\bar{t}$) with
$f_{xy}=1$ for $M_{DM}=150$ GeV. It is evident that annihilations into bosons and quarks yield too soft a spectrum, 
while annihilations into leptons are easily compatible with the data. The corresponding $\chi^2$ values are listed 
in Table~\ref{table3}. We have not displayed results for the Max set because it gives spectra similar to those for the Med set. 
Although the lowest $\chi^2$ per degree of freedom is 2 for the
Min set, we do not reject this parameter set given that uncertainties in the positron background have not yet been estimated.  
Since the Med set has a larger diffusion zone height $2L$ than the Min set, the flux of positrons incident at PAMELA is larger, 
thus requiring a smaller $B$. This explains the mode-by-mode lower boost factors for the Med set in Table~\ref{table3}.
On the other hand, the Min set has a smaller spectral index $\delta$ with relatively weaker diffusion at higher energy, 
resulting in a harder positron spectrum. This is evident from a comparison of the spectra in Figs.~\ref{fig:2} and~\ref{fig:3}.

\begin{table}[t]
\begin{tabular}{c|cc|cc}
\hline
& \ \ \ \ \ \  \ \ \ \ Med  &    &  \ \ \ \ \ \ \ \ \ \ Min  &        \\
& $B$&$\chi^2$& $B$&$\chi^2$ \\
\hline
$e^+ e^-$&30.7 & 5.63& 71.7 & 94.6 \\
$\mu^+ \mu^-$& 40.2&5.63 & 80.2 & 16.2\\
$\tau^+ \tau^-$&73.0 &32.2 & 134.6& 12.0\\
$W^+ W^-$& 119.9&31.7 & 223.6 & 15.2\\
$Z Z$& 155.7&42.6 & 277.9 & 26.9 \\
$h h$& 169.0&95.4 & 258.2 & 80.1 \\
$c \bar c$& 135.7&116.3 & 196.6 & 104.1\\
$b \bar b$& 139.7 &90.7 & 215.3 & 76.1\\
$t \bar t$& $-$ &$-$& $-$ &$-$\\
\hline
\end{tabular}
\caption{$\chi^2$ for positron spectra from two-body annihilations of DM with mass 150~GeV 
 for the Med and Min models of cosmic 
ray propagation. Results for the Max set of parameters are similar to those for the Med set.
The number of degrees of freedom in each case is 6. }
\label{table3}
\end{table}

\begin{figure}[t]
\vskip 0.2in
\mbox{\includegraphics[width=3.3in]{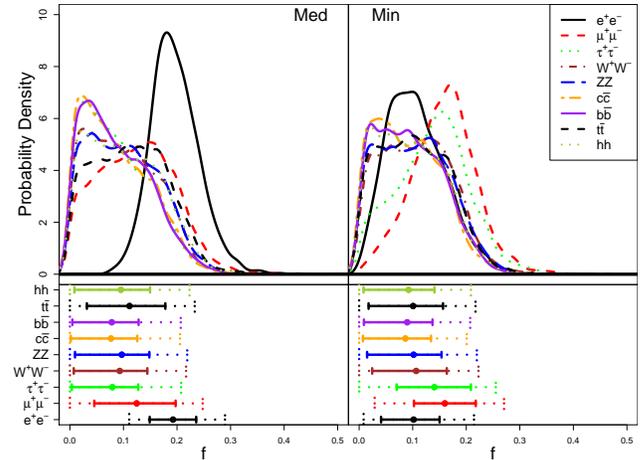}}
\caption[]{The probability distribution of the weight of each channel $f$ for the Med and Min sets. 
The medians, and $1\sigma$ and $2\sigma$ C.~L. ranges are indicated in the lower panels.
\label{fig:4}}
\end{figure}

\begin{figure}[t]
\vskip 0.2in
\mbox{\includegraphics[width=3.6in]{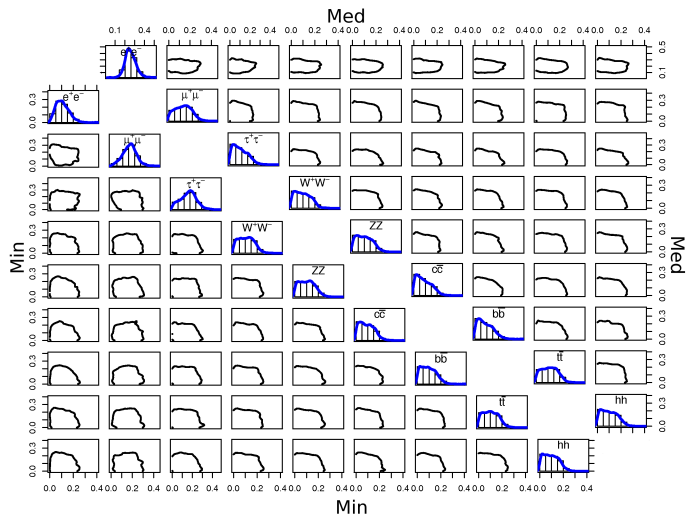}}
\caption[]{A graphical representation of the correlations between modes for the Med (upper triangle) and Min (lower triangle) 
propagation sets. The cells along the diagonal show the probability distribution of $f$ corresponding to the mode labeled.
The contour plots show the $2\sigma$ allowed regions in planes of weights, with $f$ of the column (row) mode 
along the x-axis (y-axis). No two modes are significantly correlated with each other.
\label{correl}}
\end{figure}

\begin{figure}[ht]
\vskip 0.2in
\mbox{\includegraphics[width=1.75in]{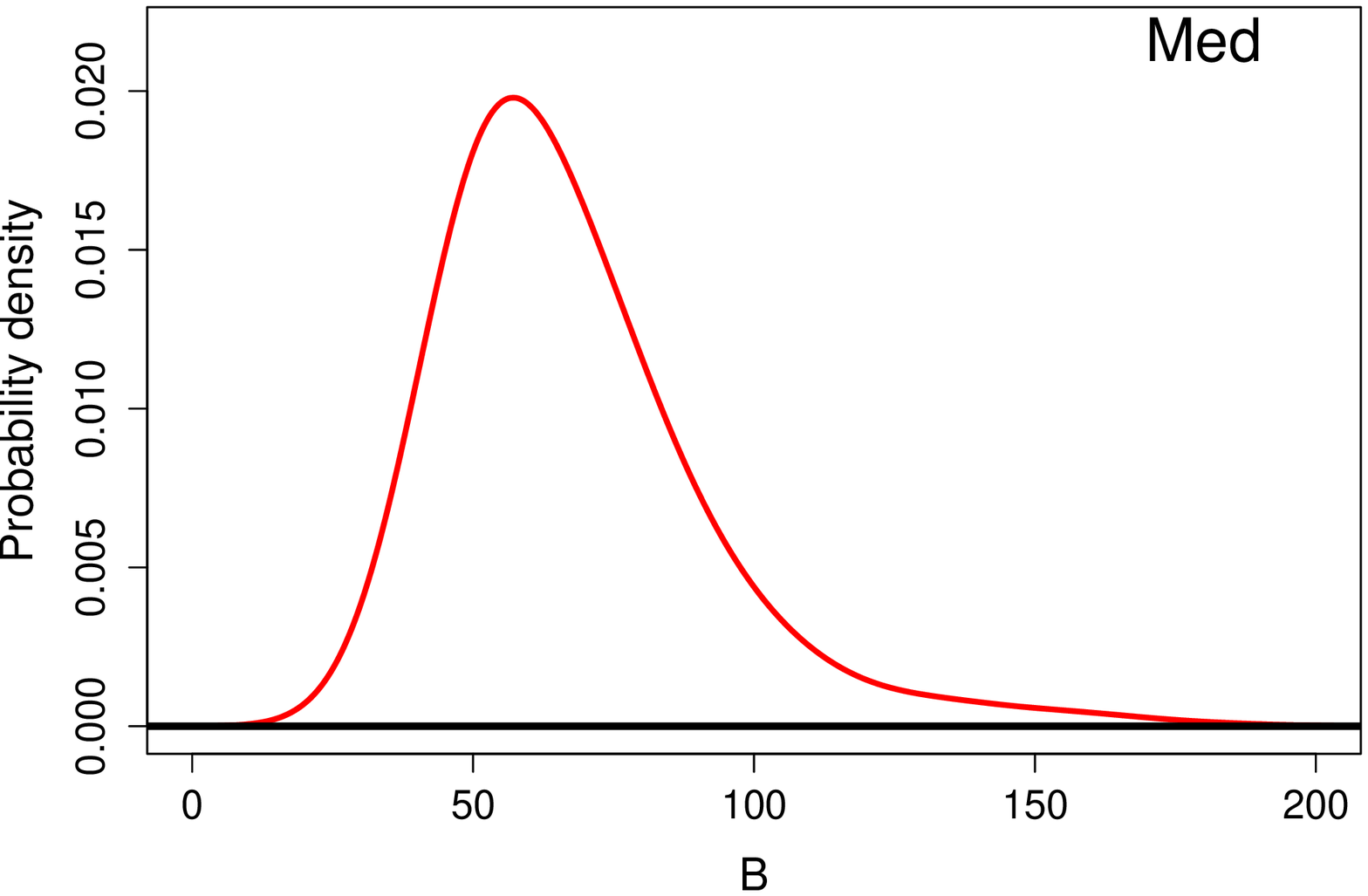}
\includegraphics[width=1.75in]{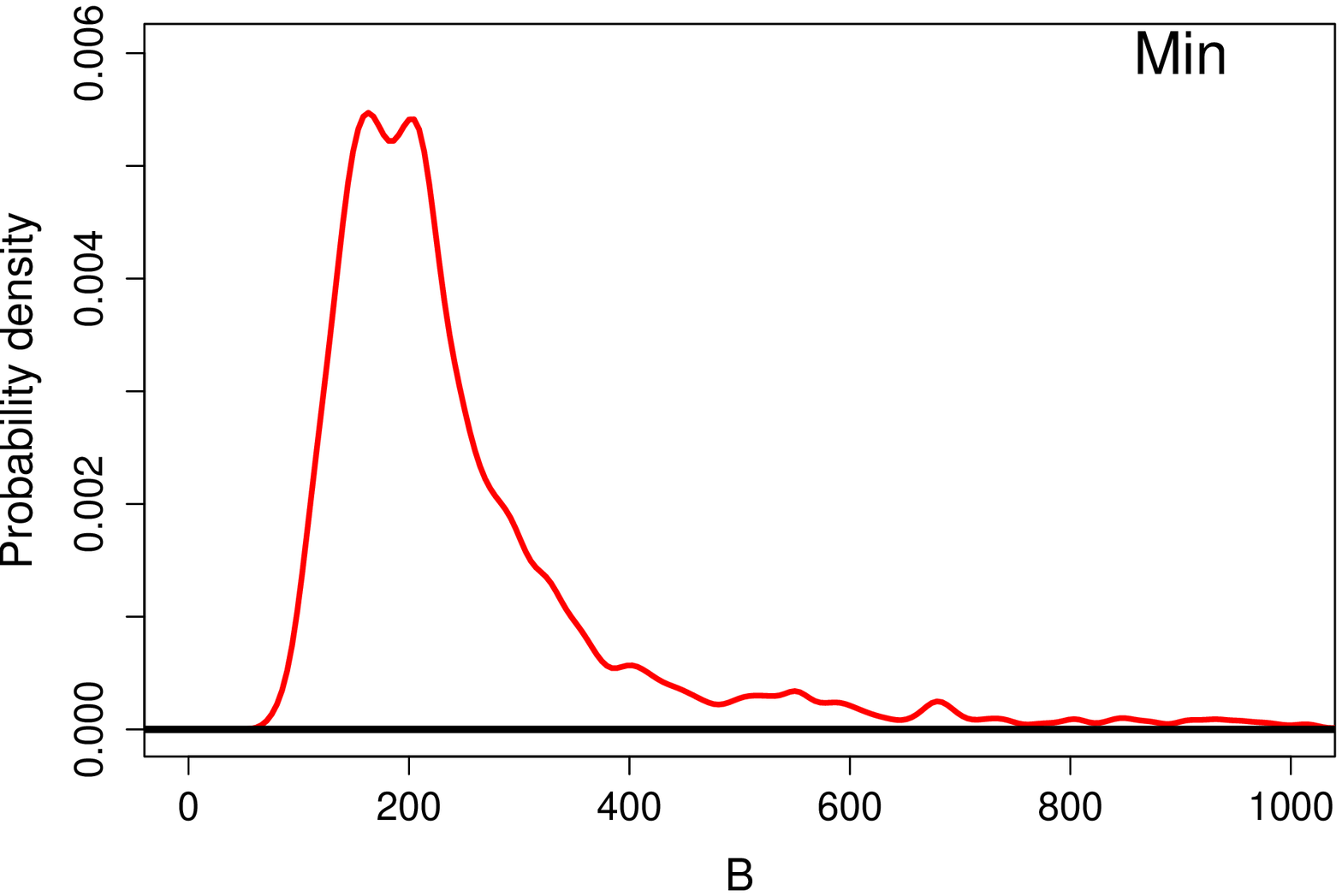}}
\caption[]{The probability distribution of the boost factor $B$ for the Med and Min propagation sets.
\label{fig:5}}
\end{figure}

We perform a Markov Chain Monte Carlo (MCMC) analysis by varying $M_{DM}$ between 100~GeV and 1~TeV, 
the boost factor $B$ and the weights $f_{xy}$
between 0 and 1 to determine
the combination of annihilation modes that fits the positron data best; see Ref.~\cite{Barger:2008qd} for a description of our
MCMC methodology. We set the annihilation cross section to be
$\langle \sigma v\rangle = 3\times 10^{-26}$ cm$^{-3}$ s$^{-1}$, which is the typical value required to reproduce  
the observed relic abundance barring co-annihilations with other particles. 
The probability distribution of $f$ is shown in Fig.~\ref{fig:4}. 
For the Med set of propagation parameters, the $e^+e^-$ mode is preferred.
For the Min set, a preference for the $\mu^+\mu^-$ and $\tau^+\tau^-$ modes is evident, and
the $e^+e^-$ mode is not favored over the non-lepton modes. In general, the data show a preference
for lepton modes. The correlation matrix for the nine modes is shown graphically in Fig.~\ref{correl}. The 
$2\sigma$ contours in planes of weights
for pairs of modes are plotted after marginalizing over all other modes.
There is essentially no correlation between modes. It is noteworthy that while the $2\sigma$ region for the 
$\mu^+\mu^-$ mode and any mode (other than $e^+e^-$) is 
consistent with $(0,0)$ for the Med set, it is not so for the Min set. This is because a soft component is necessary to
fit the data for the Min set.
The probability distribution of $B$ is shown in Fig.~\ref{fig:5}. For the WIMP annihilation
cross section we have adopted, the boost factor is about 50 for the Med set, which is not grossly unreasonable.

\begin{figure}[t]
\vskip 0.2in
\mbox{\includegraphics[width=3.3in]{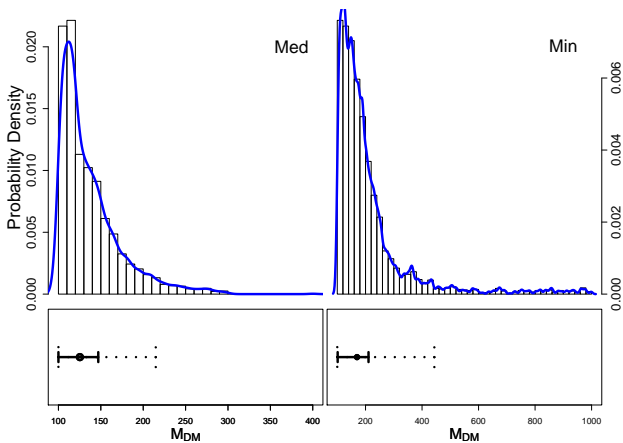}}
\caption[]{The probability distribution of $M_{DM}$ for the Med and Min sets. 
The medians which are 125~GeV (Med) and 170~GeV (Min), and $1\sigma$ and $2\sigma$ C.~L. ranges are 
indicated in the lower panels. 
\label{fig:6}}
\end{figure}

\begin{figure}[ht]
\centering\leavevmode                                                                                                                        \vskip 0.2in
\mbox{\includegraphics[width=1.75in]{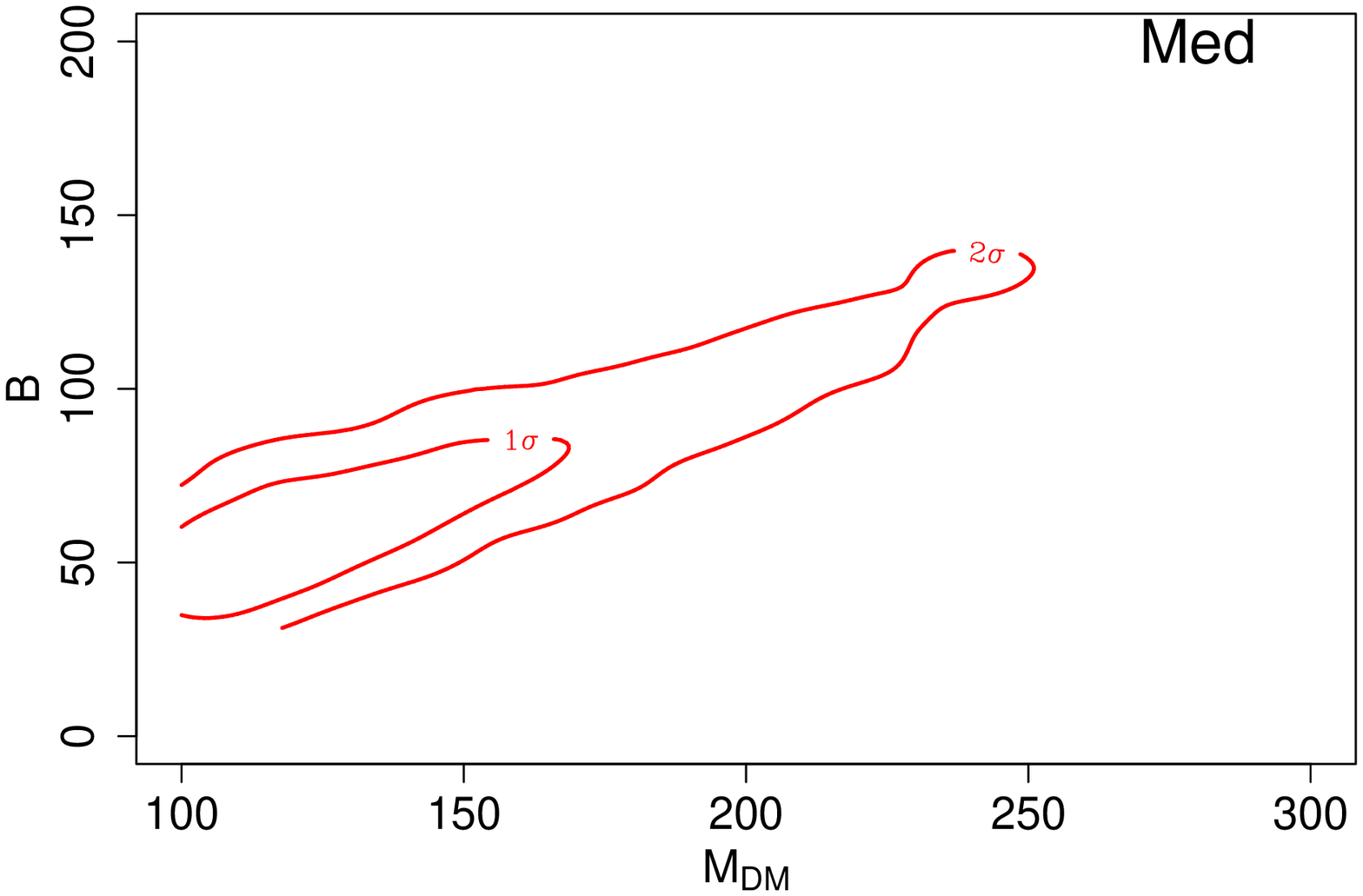}
\includegraphics[width=1.75in]{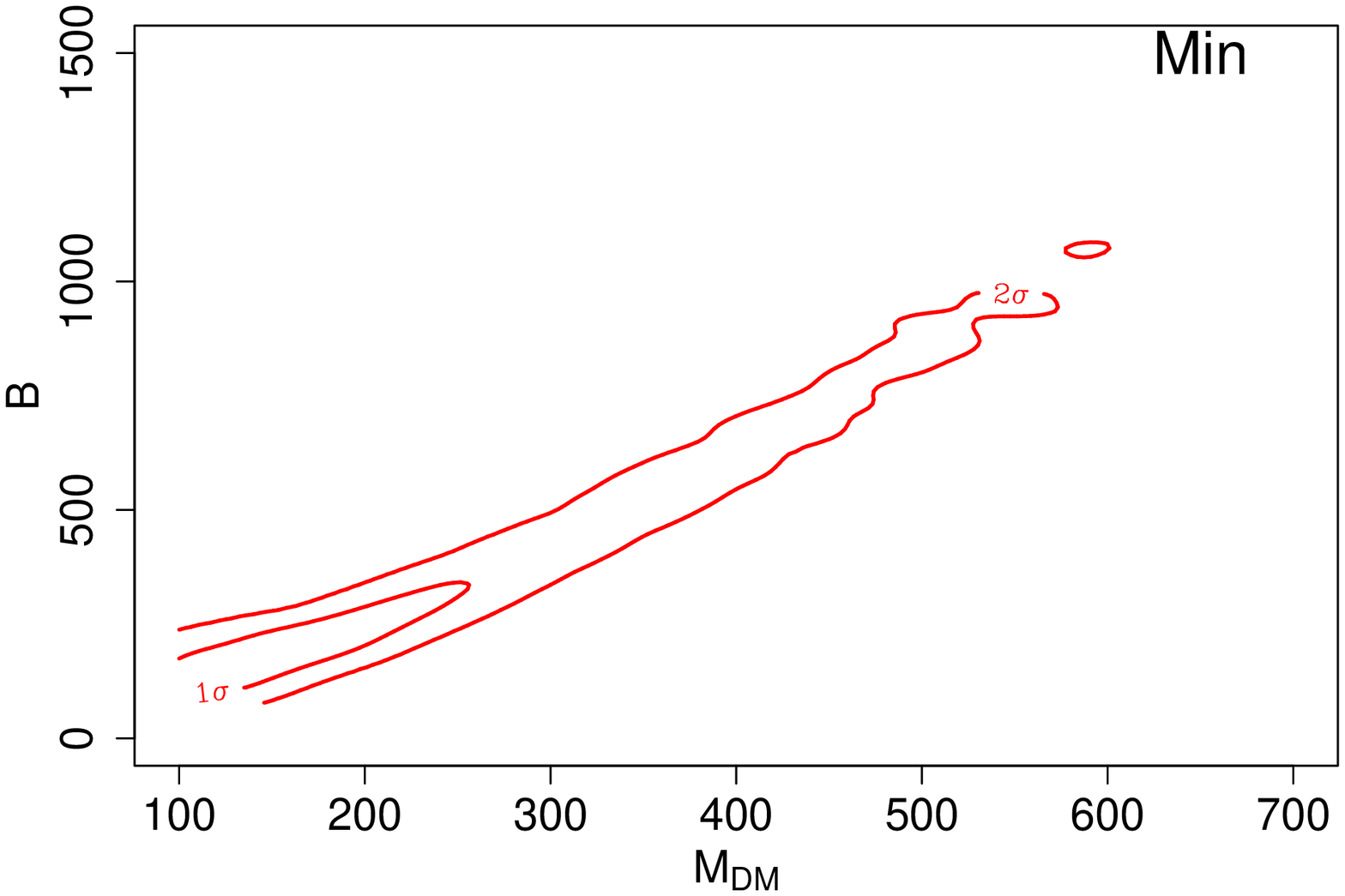}}
\caption[]{The correlation between $B$ and $M_{DM}$ for the Med and Min propagation sets.
\label{fig:7}}
\end{figure}

From Fig.~\ref{fig:6}, we see that the range of DM masses favored by the positron data depends on the details of cosmic
ray propagation. At $2\sigma$, $M_{DM}$ is below 215~GeV for the Med set and below 445~GeV for the Min set. The correlation
between $B$ and $M_{DM}$ in Fig.~\ref{fig:7} shows that ligher DM particles require a smaller boost factor to explain the 
PAMELA positron excess. Also, with the Med set of propagation parameters, very large boost factors are avoided.

\begin{figure}[t]
\vskip 0.2in
\mbox{\includegraphics[width=3.25in]{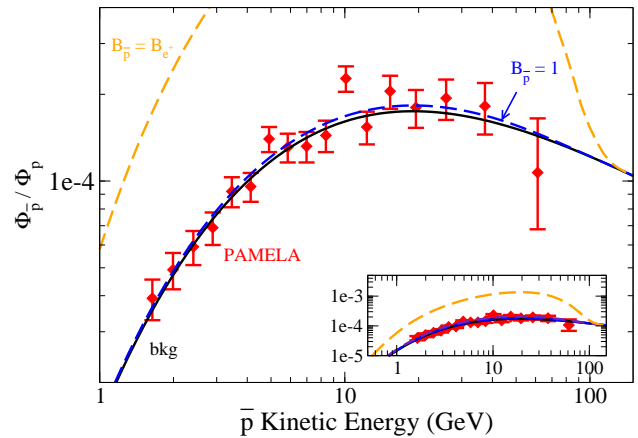}}
\caption[]{The $\bar{p}/p$ flux ratio measured by PAMELA is consistent with the expected background (solid). The dark dashed curve
is the expected spectrum from DM annihilations to $W^+W^-$, allowing for a boost factor (equal to $1$) 
that is different from that for positrons.
The light dashed curve shows the spectrum if the boost factor for the positron fraction is applied to antiprotons; the inset
shows a magnified view. 
\label{fig:8}}
\end{figure}

The antiproton spectrum measured by PAMELA up to 100 GeV~\cite{Adriani:2008zq}
 shows no deviation from the expected background~\cite{Bringmann:2006im} (which
has larger uncertainties than the positron background associated in part with the considerably greater 
propagation distance of antiprotons).
Since our approach
is model-independent we can not make definite statements about consistency with the cosmic antiproton data.
By choosing an appropriate
boost factor (which can be different from that for positrons) and 
appropriately modelling the propagation of antiprotons, it is easy to remain in agreement
with the data. Within our approach it is also possible to have consistency
by suppressing the DM annihilation branching fraction to antiquarks. As an illustration, in Fig.~\ref{fig:8} we show 
the antiproton to proton flux ratio measured by PAMELA, and the theoretical expectation for the $W^+W^-$ channel from
annihilations of DM of mass 150~GeV. Boost factors for the antiproton flux below 3.3 yield agreement at the 2$\sigma$ C.~L.  
The light dashed curve shows the $\bar{p}/p$ flux ratio if the antiproton boost factor is taken to be equal to the positron
boost factor that fits the positron spectrum. Clearly, different boost factors are necessary. The 1-2 orders of magnitude
difference in the $e^+$ and $\bar{p}$ boost factors is a problem.

\textbf{Conclusions.} Our results are summarized in the figures. We have shown 
that the PAMELA positron excess does not favor a 
DM particle of a particular spin. The data
do not discriminate between positron spectra from direct production and from secondary decays of polarized $W$ bosons. 
However, PAMELA is expected to collect positrons up to about 270~GeV.
With those data it should be possible to draw stronger conclusions. If the data show a line, popular SUSY 
models will be in danger of being excluded and
models with extra dimensions and collective symmetry breaking will gain support since they have spin-1 DM. 
Models with Dirac fermions as DM will also be viable. On the other hand if the data
roll-over smoothly near the endpoint, and are fit well by positrons from transversely polarized $W$ bosons, SUSY will be indicated. If
positrons from longitudinally polarized $W$ bosons are preferred by the data, neutralino DM will be in jeopardy, and
the DM candidates of mUED and LHT will be preferred. To make such fine distinctions in spectral shapes will require much 
larger datasets from PAMELA and the Alpha Magnetic Spectrometer.

By considering nine different two-body annihilation channels with arbitrary weights, for dark matter lighter than 1~TeV, 
we found that lepton modes are generally
preferred by the positron data, and which lepton modes are favored depends on the details of cosmic ray propagation. The 
$\mu^+\mu^-$ and $\tau^+\tau^-$ modes fit the data better than the obvious $e^+e^-$ mode for the Min set. Also, we found that  
dark matter masses selected by the data depend on the propagation model. The $2\sigma$ upper limit
 is 215~GeV for the Med set of propagation parameters and 445~GeV 
for the Min set. Results for the Max set are similar to those for the Mid set.

It is important to bear in mind that although astrophysical processes are
expected to produce a positron background that falls with energy, it may still be that astrophysical
sources such as pulsars could mimic the putative DM signal.
Confidence in the DM interpretation will be strengthened by
 signals in other experiments, involving 
both direct and indirect detection methods.


{\it Acknowledgments.}
This research was supported
by DOE Grant Nos.~DE-FG02-04ER41308,  DE-FG02-95ER40896, DE-FG02-84ER40173 and DE-AC02-06CH11357,
by NSF Grant No.~PHY-0544278, and by the Wisconsin Alumni Research Foundation.


\end{document}